\documentclass{llncs}

\usepackage{booktabs} 

\usepackage{algorithm}
\usepackage{algorithmic}
\usepackage{color}
\usepackage{amsmath}
\usepackage{amssymb}
\usepackage{graphicx}     
\usepackage{xspace}     

\usepackage[numbers,sort]{natbib}

\newcommand\kha{\texttt{Kharita}\xspace}

\newcommand\mapfuse{\texttt{MapFuse}\xspace}

\newcommand{\bigO}{\ensuremath{\mathcal{O}\xspace}}

\title{MapFuse: Road Network Fusion for\\Incremental Map Updates}

\begin{document}

\author{Rade Stanojevic$^1$, Sofiane Abbar$^1$, Saravanan Thirumuruganathan$^1$, Gianmarco De Francisci Morales$^1$, Sanjay Chawla$^1$, Fethi Filali$^2$, Ahid Aleimat$^2$}
  \institute{$^1$Qatar Computing Research Institute, HBKU\\
  P.O. Box 5825, Doha\\
   $^2$Qatar Mobility Innovation Center, QSTP\\
   P.O. Box 210531, Doha\\
 \email{\{rstanojevic, sabbar, sthirumuruganathan, gmorales, schawla\}@hbku.edu.qa, \{filali,ahide\}@qmic.com}
   }

\maketitle

\begin{abstract}
In the recent years a number of novel, automatic map-inference techniques have been proposed, which derive road-network from a cohort of GPS traces collected by a fleet of vehicles. In spite of considerable attention, these maps are imperfect in many ways: they create an abundance of spurious connections, have poor coverage, and are visually confusing. Hence, commercial and crowd-sourced mapping services heavily use human annotation to minimize the mapping errors.
Consequently, their response to changes in the road network is inevitably slow.

In this paper we describe \mapfuse, a system which fuses a human-annotated map (e.g., OpenStreetMap) with any automatically inferred map, thus effectively enabling quick map updates. In addition to new road creation, we study in depth road closure, which have not been examined in the past. By leveraging solid, human-annotated maps with minor corrections, we derive maps which minimize the trajectory matching errors due to both road network change and imperfect map inference of fully-automatic approaches.

\end{abstract}

\section{Introduction}
\label{sec:introduction}

\noindent {\bf Map Fusion Problem:}
Generating accurate maps from geospatial data is an active area of research.
A number of these works~\cite{biagioni2012inferring,cao2009gps,chen2016city,edelkamp2003route} utilize crowd-sourced GPS data, e.g., from smartphones.
An alternate strain of work tries to use other sources such as satellite images \cite{mnih2010learning}.
Despite considerable interest and effort by the research community, the existing automatic map inference solutions have a number of shortcomings, including: limited coverage,  visually confusing layout, spurious roads, and imperfect turn restrictions. 
Hence, commercial maps such as Google Maps, Nokia HERE, and Apple Maps often use multiple sources of data information to generate initial maps,
and then rely heavily on humans (both annotators and volunteers) to detect and correct the possible imperfections.
However, the involvement of humans results in a very slow response in updating maps when a change in the road network occurs.
In many cities in Asia and Africa, which are under heavy construction, this process results in substantial latency.
One potential way to solve this issue is to automatically update the map using GPS traces given an existing map.
However, most of those approaches are simple adaptations of classical map inference algorithms and suffer from the same disadvantages.
In this work, we advocate for a new approach - Map Fusion - which automatically fuses two maps.
One of the maps is a high-quality slowly updated map such as OpenStreetMap (OSM) \cite{openstreetmap} or Google Maps \cite{googlemaps},
while the other one is an automatically inferred map with incomplete coverage and imperfect topological structure.
Our proposed system, \mapfuse, synthesizes a new map that overcomes the deficiencies of the two maps discussed above.
In the rest of the section, we enunciate this overall approach.

\subsection{Challenges in Fully Automatic Map Inference}
As mentioned above, there has been extensive work (see surveys~\cite{biagioni2012inferring,ahmed2015comparison,liu2012mining})
on automatic map creation from GPS traces.
However, these algorithms - both academic and commercial -  face a number of important challenges.
We now highlight three of the major ones.

\begin{itemize}
    \item {\it Poor coverage}. The popularity of roads segments in the road network (measured, say, in number of trajectories which pass by the segment) is very skewed.
    While a few road segments (e.g., those lying on a highway) carry a massive number of trajectories, a large fraction of roads serves only a handful of cars. Hence, a vehicle fleet which opportunistically collects the GPS data needs to collect a massive amount of spatial samples in order to have a decent coverage of the road network.
    In the case of the fleet whose data we analyzed in this work, if we denote by $L$ the total length of all the roads in Doha ($L$ is in the order of 10s of thousands of kilometers) our data, which corresponds to the trajectories with overall length of $175 \cdot L$, covers only about 48\% of the road segments (see Figure \ref{fig:coverage}). In order to cover close to 100\% of the road network with such opportunistic GPS probes, one would need to collect from one to two orders of magnitude more
        data, which in case of Doha would translate to 10s or 100s of millions of kilometers of driving. Thus, independent of the map-inference method one utilizes, one needs to have an extremely high-volume of opportunistically collected GPS data in order to cover large portions of the road network. 

\begin{figure}[ht]
\centering
 \includegraphics[width=1\textwidth]{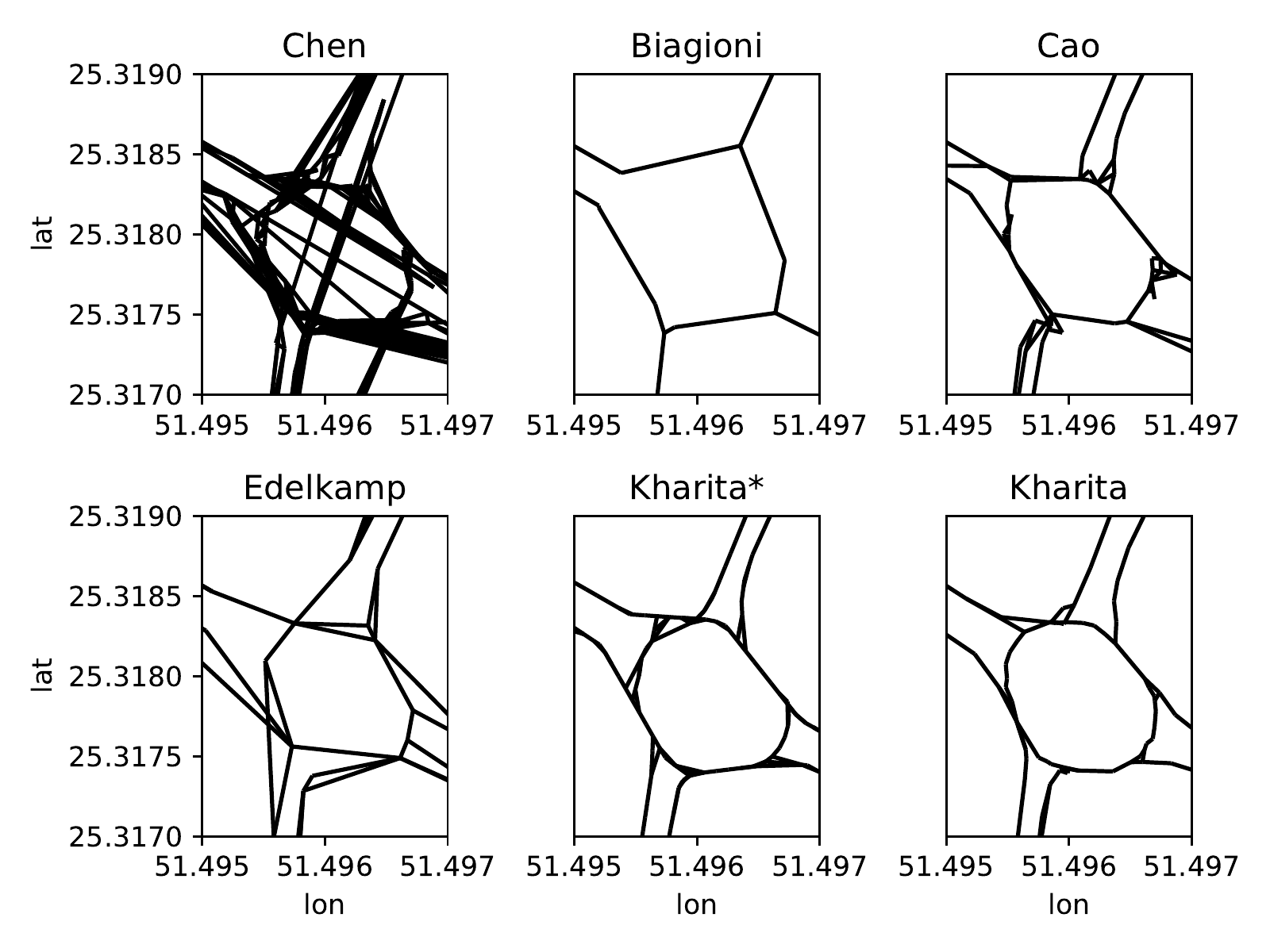}
  \caption{Automatically inferred maps of 6 existing methods.}
  \label{fig:sixtv}      
\end{figure}

\item {\it Visually confusing outlook}. Most of the existing approaches do not control for the visual appearance of their maps, and hence the resulting maps have rather confusing look and are not visually appealing. In Figure \ref{fig:sixtv} we depict maps of a prominent ``TV roundabout'' in Doha derived by several well-known map-inference algorithms \cite{biagioni2012inferring,cao2009gps,chen2016city,edelkamp2003route,stanojevic2017kharita}. Due to different nature of their inference process,
    they all have some unique features, yet they all have spurious or missing road segments, which can confuse the end-user and the navigation system which may utilize such maps.

\item  {\it Low topological accuracy}. Possibly the most serious concern regarding the existing map-inference methods is their low topological accuracy. Namely, due to the GPS noise as well as the inability to efficiently handle such noise, all existing methods often miss the connections between road segments or infer non-existing connections between road segments. Such topological inaccuracies are absolutely non-tolerable, yet existing solutions have topological Biagioni $F1$-score\footnote{Biagioni $F1$-score is a well known metric for measuring the topological accuracy of a map and lies in the range $[0,1]$ with 0 being absolutely wrong map, and 1 being a perfect map.} \cite{biagioni2012inferring} in the range of 0.6 to 0.8~\cite{BiagioniE12,stanojevic2017kharita}. We believe that a commercially acceptable map would likely need to have Biagioni $F1$-scores in the nearest proximity of 1. 
\end{itemize}

\subsection{Challenges for Automatic Map Updates}
TomTom reports that 15\% of roads change each year in some way~\cite{wang2013crowdatlas}. The road changes are particularly common in many developing countries in Asia and Africa due to rapid construction of new roads. For example, thousands of kilometers of new expressways have been constructed each year in China and India for the past few years~\cite{wang2017hymu}. Automating the map update in a way that minimizes the disruption to the original map is of paramount importance. 
There has been extensive work on automatically updating an existing map using newly acquired GPS data (see Section~\ref{sec:relatedWork} for details).
However, many of these algorithms are often simple adaptations of existing batch map-inference algorithms, and suffer from the same issues mentioned above.
In addition, they often start with an automatically generated map which also suffers from the issues mentioned above.
Hence the resulting map is often of substandard quality.

\subsection{Challenges for Hybrid Map Updates}

\begin{figure}
\centering
 \includegraphics[width=0.7\textwidth]{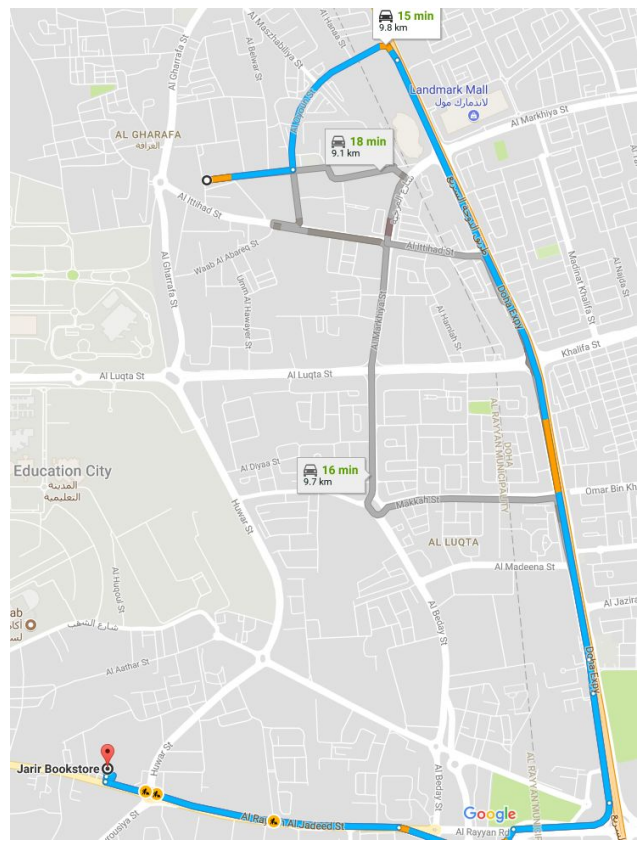}
  \caption{Google maps route suggestion between two locations in Doha are almost twice longer (in length and duration) than the optimal route.}
  \label{fig:gharaffa}      
\end{figure}

According to the discussion so far, we believe that a hybrid method involving automatic algorithms along with humans is the way forward.
The substandard quality of maps from purely automated means is often unacceptable for commercial map systems such as Google Maps, Apple Maps, Bing Maps, Nokia HERE, and Tom Tom.
The creation of these maps is in many ways automated, however it requires human attention to examine possible places of interest. For example, Google Maps has a large team of so called operators who ensure the validity and consistency  of the Google maps~\cite{elbow} and hence any possible change in the road network needs to be approved by one of
the operators. Similarly, the largest global crowd-sourced mapping effort OpenStreetMap (OSM) updates around 1M nodes per day. These maps have reasonably high accuracy in most cities with static road infrastructure. 

However, even this approach has some fundamental limitations.
Due to the human in the loop, they suffer from slow update response when changes happen (see Figure \ref{fig:gharaffa}). 
In many cities such as Doha, there are constant and large changes in road networks, that are not reflected in the maps in a timely manner.
Conversely, automated algorithms often ignore the fact that most urban areas globally already have a fairly accurate map infrastructure. Not utilizing such great resource to construct the map (as most automatic map inference solutions do) is unfortunate and hurts the overall map inference process.
Let us illustrate this effect with a real-world example.

In the city like Doha, with a very dynamic road network,\footnote{Influenced by a rapid construction of the city metro and a number of ongoing infrastructure projects.} the quality of existing maps is rather poor. For example, when one queries Google Maps for a route suggestion between two points in west Doha (see Figure \ref{fig:gharaffa}), the suggested routes are almost twice as long (in both time and length) than the optimal one.  Even though the optimal route has existed for over a year, the Google Maps has not yet updated the relevant portion of the map to reflect the current layout.

\subsection{Proposed Approach}
In this paper we propose \mapfuse, a system for map fusion which automatically merges two maps. Specifically, we seek to fuse (1) a high-quality slowly-updated map such as OSM \cite{openstreetmap} or Google Maps \cite{googlemaps} and (2) an automatically-inferred one, with incomplete coverage and imperfect topological structure.
\mapfuse produces a map which overcomes the deficiencies of the two maps discussed above. 

In contrast with the existing approaches on map updating, which update the existing map (say OSM) by using a set of GPS trajectories via a specific map-inference tool, \mapfuse is oblivious to the map inference approach one wishes to use to capture the road network segments and the interconnections between them. Hence we can fuse \emph{any} map to the existing underlying map. This is important because existing map inference solutions suffer from a number of issues, and future solutions will most certainly rectify many of those. Fusing such better-inferred maps will most certainly lead to higher quality maps.

Finally, a very relevant aspect of map updating are road closures (both temporary and permanent) which are overlooked by the previous work on map updating, as it focuses only on new road additions~\cite{shan2015cobweb,wang2013crowdatlas}. We use the GPS trajectory data to understand the road dynamics and infer road closures as soon as they happen. 

\textbf{Summary of Contributions:}
\vspace{-0.5\baselineskip}
\begin{itemize}
    \item We introduce the problem of map fusion, which seeks to update a base map with another inferred map, as a geometric graph matching problem and show it can be treated as a minimal vertex cover problem on an appropriately-defined bipartite graph. 
    \item Due to the size of the graphs representing the two maps (which can have hundreds of thousands of nodes) the polynomial solution to the bipartite vertex cover problem is not practical and we propose an efficient heuristic that fuses two maps.  
    \item We suggest a new methodology for inferring closed road segments which utilizes dynamic statistics of the roads as well as a node centrality measure. As an unexpected advantage of our closure detection we identify the errors in the OSM maps (e.g., we can automatically pinpoint several roundabouts which are represented in the OSM as two-way roads, while they are obviously one-way only) which can be harmful to the navigation systems.
    \item Using a set of GPS trajectories from a fleet of vehicles in Doha we demonstrate that the fused map is more accurate than either of the two maps, and reduces the average/median/99th-percentile trajectory matching error by 30\%.
\end{itemize}
   
\section{Related Work}
\label{sec:relatedWork}
       
\noindent {\bf Map Inference:}
Constructing maps from crowdsourced GPS traces has been extensively studied 
(see surveys~\cite{biagioni2012inferring,ahmed2015comparison,liu2012mining}.
K-Means based algorithms cluster the GPS points and link the resulting clusters into a routable map.
Representative works include~\cite{edelkamp2003route,agamennoni2011robust,schroedl2004mining}.
Kernel density estimation (KDE) based algorithms such as \cite{chen2008roads,davies2006scalable,shi2009automatic}
transform the GPS points into a density discretized image that are processed by image processing techniques to obtain maps. 
Trace merging based approaches start with an empty map and carefully add traces into it. 
Representative works include \cite{cao2009gps,ahmed2012constructing}.

\noindent {\bf Maintaining Maps:}
Maintaining maps is closely related to map inference and often the algorithms for map maintenance are adaptations of those for map inference.
Nevertheless, there are some subtle differences.
While one can indeed obtain an updated map by re-running the entire inference pipeline, it is often efficient - in terms of both time and data - 
to treat it as a separate problem.

Recall that almost 15\% of roads change every year in the US~\cite{wang2013crowdatlas}.
This number is even higher in many developing countries in Asia and Africa due to rapid construction of new roads.
For example, thousands of kilometers of new expressways are being constructed each year in China and India for the past few years~\cite{wang2017hymu}.
This necessitates research into work that maintain and update maps as and when new GPS data points arrive.
Some representative work include~\cite{ahmed2012constructing,schroedl2004mining,vanall,bruntrup2005incremental,wang2013crowdatlas,zhang2010integration,shan2015cobweb,wu2015glue,wang2017hymu}.
However, most of these approaches do not have good practical performance and are very sensitive to differential sampling rates, disparity in data points, GPS errors etc.
Often, these algorithms seek to directly extend one of the three approaches and suffer from bottlenecks arising from algorithmic step that is fundamental to it (such as clustering, density estimation, clarification, map matching) etc.

Additionally, while most of the prior work handle the simple case of new road additions, road closures are rarely addressed. 
CrowdAtlas~\cite{wang2013crowdatlas} is exception that uses a simple heuristic in which each road segment is assigned an appropriate timeout proportional (3x) to the  maximum time observed between the traversal of two successive vehicles in a training window. To cope with the cold start problem, no timeout is set for a segment until it has accumulated at least a week of data and at least five traces. Thus, most residential roads have no timeout established.

\noindent {\bf Graph Matching:}
Given two graphs, identifying if one graph is a subgraph of another is known to be NP-Complete~\cite{garey2002computers}.
In fact, even identifying the minimal set of `edits' to transform one graph to another is also NP-Complete~ \cite{zeng2009comparing}.
However, it is possible to apply a number of heuristics for the case of road networks to solve this problem effectively.
Matching of two road networks has been extensively studied due to its practical importance.
The process of integrating different geospatial data to get new cartographic products is called map conflation.
See~\cite{ruiz2011digital} for a review of techniques used. 
Often, a wide variety of information including spatial features (such as distances, angles, shapes of the map) 
and topographical information (such as neighborhood) are used.
For example, \cite{yang2013probabilistic} proposed a heuristic probabilistic relaxation procedure to integrate multi-source geospatial data
by using similarities between shapes.
Recently, \cite{du2015tool} studied the problem of integrating authoritative geo-spatial data (such as OpenStreetMap) with crowdsourced GPS information.
However, they use auxiliary information such as names and types of POIs that may not always be available.

\section{Problem Formulation}
\label{sec:problem}
A common representation of a map in the map-inference literature is a directed graph as following. A map is a geometric graph $G(V,E,L)$, where $V$ is the set of vertices, $E \subseteq V \times V$ is the set of edges connecting pairs of vertices, and $L : V \rightarrow \mathbb{R}^2$ is a location function which assigns coordinates (latitude and longitude) to each vertex.
  
Given two instances of such graphs (maps), $G_1$ and $G_2$, our goal is to create a new fused graph $G_f = f(G_1, G_2)$ which preserves some properties of the source graphs.
In particular, we wish for the connectivity of the fused graph to subsume the connectivity of the source graphs.
However, we also wish to do so with the minimum number of edges, in order to avoid unnecessary and spurious ones.

In order to express the connectivity property, we consider the set of shortest paths $\pi_i$ within each graph $G_i$.
The fused graph $G_f$ should be so that
\begin{equation}\label{eq:mapfusedef}
 \forall p \in \pi_i ,\, \exists \hat{p} \in \pi_f \, \text{ s.t. } d(p, \hat{p}) \leq \theta ,\; i \in \{1,2\},
\end{equation}
where $d(\cdot,\cdot)$ is a suitable distance function between paths which takes into account their geometry, and $\theta$ is a user-specified tolerance parameter.
In our paper we use the following distance function
\[
d(p_0,p_1) = \min_{i=0,1} \max_{u\in p_i} v(u,p_{1-i})
\]
where $v(u,p)$ is the minimum distance between a point $u$ and path $p$ measured in meters. Thus a small $d(p_0,p_1)$ indicates that one of the two paths can be matched onto the other. 

In addition, we wish to find the ``minimum'' such graph, i.e., the one that minimizes the sum of the lengths of its shortest paths:
\[
\arg\min_{G_f} \sum_{p \in \pi_f} \ell(p).
\]

This problem formulation can be reconducted to a minimum vertex cover problem on a suitably-defined bipartite graph $H(\pi_1, \pi_2, F)$.
The two sets of vertices in $H$ are all the possible shortest paths in $G_1$ and in $G_2$ ($\pi_1$ and $\pi_2$, respectively).
There is an edge $(u,v)$ between two elements $u$ and $v$ if their distance is below the threshold, i.e.
\[
(u,v) \in F \iff d(u,v) \leq \theta ,\, u \in \pi_1, v \in \pi_2.
\]

Finding a minimum vertex cover $M$ on $H$ is equivalent to finding a minimum set of shortest paths such that their union maintains the connectivity property of the two source graphs.
Therefore, $G_f$ can be build from the union of these paths $M \subseteq \pi_1 \cup \pi_2$.

Note that due to K\"{o}nig's theorem, the minimum vertex cover problem on a bipartite graph is actually tractable in polynomial time (and not NP-hard as in the general case).
However, the size of the problem is $\bigO(n^2)$, and that to materialize $H$ na\"{i}vely we need to compute $\bigO(n^4)$ distances between pairs of shortest paths.

Graphs representing the OSM and inferred maps in a large city such as Doha have more than $n = \bigO(100K)$ nodes. Hence the polynomial solution we hinted above is impractical.
Therefore in the following section we propose a simple and efficient heuristic for tackling map fusion problem. 

\section{New Roads Detection}
\label{sec:addition}

A common approach used in the literature~\cite{shan2015cobweb,wang2013crowdatlas} to identify or detect new roads is the following. First,  run a map matching algorithm between an existing map and a collection of GPS trajectories to identify the subset of trajectories that remain unmatched. Second, run some road creation algorithm on the collection of unmatched trajectories to identify the new roads. Finally, link the newly created road segments to the existing map. That is, at the heart of the process, an algorithm is required to create roads from GPS points, which is exactly what all map inference algorithms do. Thus, it is hard to understand the real added value of map updating algorithms compared to what map inference algorithms do. For instance, if we assume that the initial map is very sparse, then it becomes clear that map update algorithms will be creating most of the road network, just like map inference algorithms do. 
Another way to look at the issue is to consider an initial empty map: in this case the map update and map inference become equivalent problems. 

In our work, we take a slightly different approach. We assume that two maps are given to us. One that represents the base map (e.g., OSM) and another one that is generated using GPS traces via one of the many map inference algorithms available. The problem is then redefined as merging these two maps. 
    
The function $FindOutliers$  takes as input two maps $M_1$ (original) and $M_2$ (inferred), and generates a set of outliers. Outliers are set of nodes in the map $M_2$ which are at distance at least $\theta$ (here we use $\theta = 20m$). Mappings link nodes in $M_2$ to $M_1$, whereas outliers are those nodes in $M_2$ that have no correspondents in $M_1$. These nodes are considered as candidates to be part of new road segments not covered in $M_1$. Our road addition procedure (see Algorithm \ref{alg:rap}) works as follows.

\begin{algorithm}
    \caption{\mapfuse}
    \label{alg:rap}
    \begin{algorithmic}[1]
	\STATE {\bf Input:} Base road map $M_1$, inferred map $M_2$ 
	\STATE {\bf Parameters:} collision radius ($r$, in meters)
	\STATE $outliers$ = $FindOutliers(M_1,M_2)$
	\STATE $DRS$ = $Subgraph(M_2, outliers)$
	\FOR {each $o \in outliers$}
	  \STATE compute $distance(o, M_1)$
	\ENDFOR
	\STATE $outliers = Sort(outliers)$ in decreasing order of distance to $M_1$
	\FOR {each $o \in outliers$}
	  \STATE $sg = BFS(o, DRS)$
	  \FOR {each node $n \in sg$}
	    \IF {$distance(n, M_1) \leq r$}
	      \STATE $merge(n, argmin(n, M_1))$
	    \ENDIF
	    \STATE $outliers = outliers - \{n\}$
	  \ENDFOR
	\ENDFOR
        \RETURN $M_1$
    \end{algorithmic}
\end{algorithm}
In line 4, the sub-graph of newly detected roads (DRS) in $M_2$ is generated from the outliers. In lines $5-8$, the outliers are sorted in a decreasing order of their geometric distance to $M_1$.  
The intuition here is that the farther a node is from $M_1$, the more likely that node lays on a new road segment not covered by $M_1$. 
Outlier nodes are then processed in their order as follows. For each node $o$, we run a breadth first search (BFS) in $M_2$ starting from $o$ until it reaches a leaf node or a node that is within a radius $r$ (e.g., 2 meters) from $M_1$. Leaf nodes are assumed to be dead ends of newly detected road segments whereas nodes within a radius distance $r$ from $M_1$ are assumed to belong to $M_1$. Nodes in the latter case are then merged with their closest nodes in $M_1$ as per line 13. 

It is not difficult to see that the output $M_f$ of  above algorithm satisfies the condition from the Eq. (\ref{eq:mapfusedef}). All paths from $M_1$ are indeed in $M_f$ and are obviously matched by paths of $M_f$, the nodes from $M_2$ which are more than $\theta$ away from $M_1$ eventually get merged into the $M_f$ and clearly satisfy the matching requirement (\ref{eq:mapfusedef}).

\section{Closed Roads Detection}
\label{sec:deletions}
Recall that the input to our process is the original map $M_1$, GPS-level trajectory data and the automatically inferred map $M_2$. An important characteristic of the road network are road closures, which are sometimes permanent, but often temporary. Unfortunately, road closures have been overlooked by previous map-inference/map-update literature and in this section we propose two novel techniques for inferring road closures. The first one is `static', in that it infers the road closures on a fixed input of trajectory data on the roads which have been closed prior to the start of the data collection. The second technique is more dynamic, as it observes the time series of the trajectories passing by a given road segment and by looking for anomalies is such time-series it effectively detects the road closures on the segments which have previously carried some trajectories in the data.
   
\subsection{Cold-start road closure detection}
\label{sec:coldstartdetection}
As we hinted above, trajectory data collection inevitably has a starting point which is determined by either the functionality of the probe and the back-end system which stores the data, or by privacy regulations which may require  sensitive trajectory data to be deleted after a period of time elapses. 

\begin{figure}[ht]
\centering
  \includegraphics[width=0.7\textwidth]{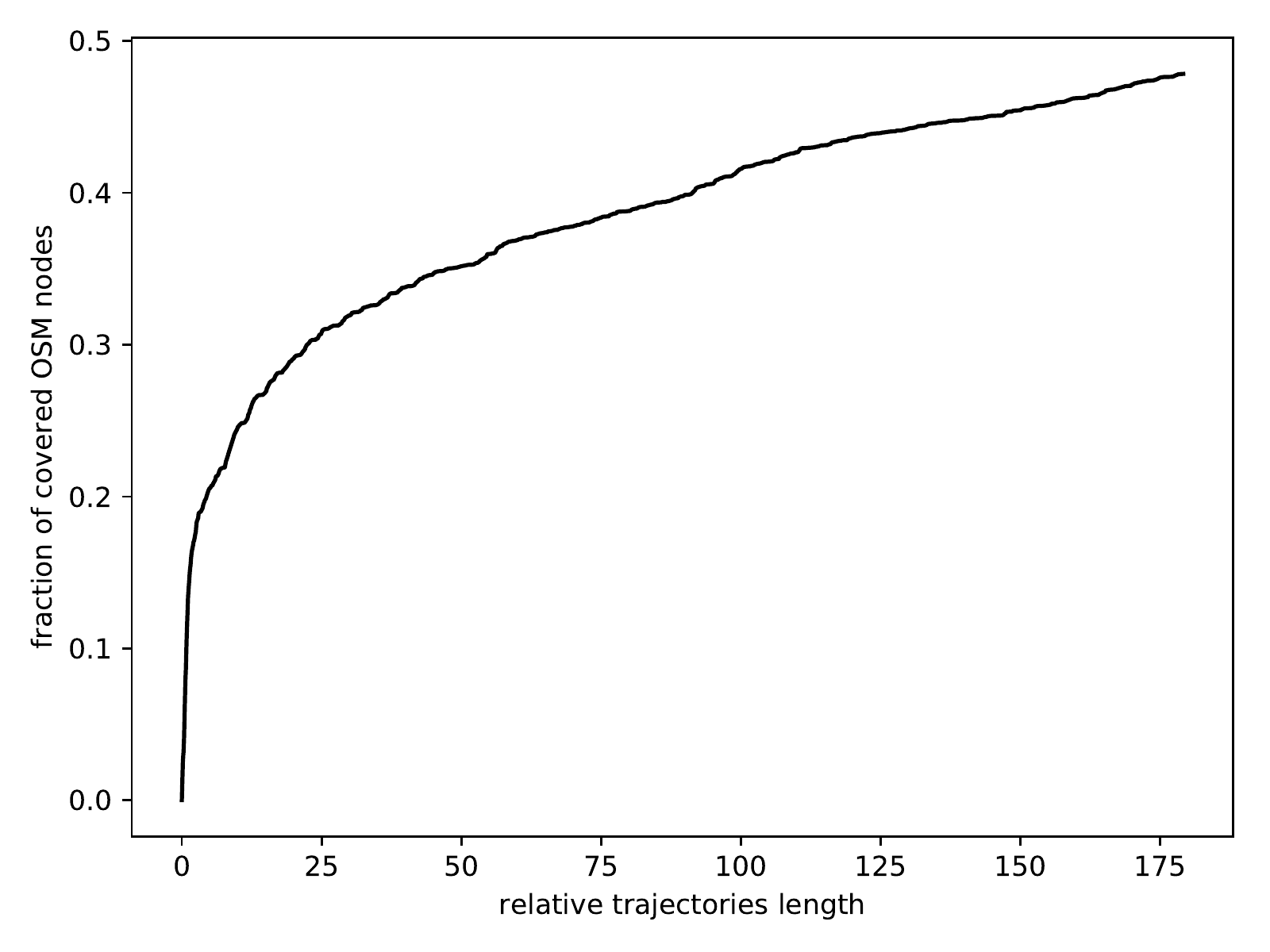}
  \caption{Fraction of OSM nodes which are covered by at least one trajectory as a function of relative trajectory length defined as the ratio between the total length of all trajectories up to a point in time and total length of the road infrastructure. For close to 100\% coverage one would need to have very, very, large trajectory dataset.}
  \label{fig:coverage}      
\end{figure}

What makes detection of closed road segments (from map $M_1$) difficult is the fact that there is a very high skew in the frequency of trajectories on different road segments: some segments (e.g., highways) carry a large number of trajectories while others in the capillary roads may not carry even a single trajectory. In Figure \ref{fig:coverage} we show how many new segments are `discovered' as more driving data is collected. If we denote by $L$ the total length of the road network, after trajectories with total length of $L$, only about 10\% of the unique road segments are touched by those trajectories. After the total trajectory length gets to $10L$ they touch around 22\% unique road segments. With all trajectories in our dataset with total length of $175L$, we get to detect only about 48\% of the road network.

\begin{figure}[ht]
\centering
  \includegraphics[width=0.7\textwidth]{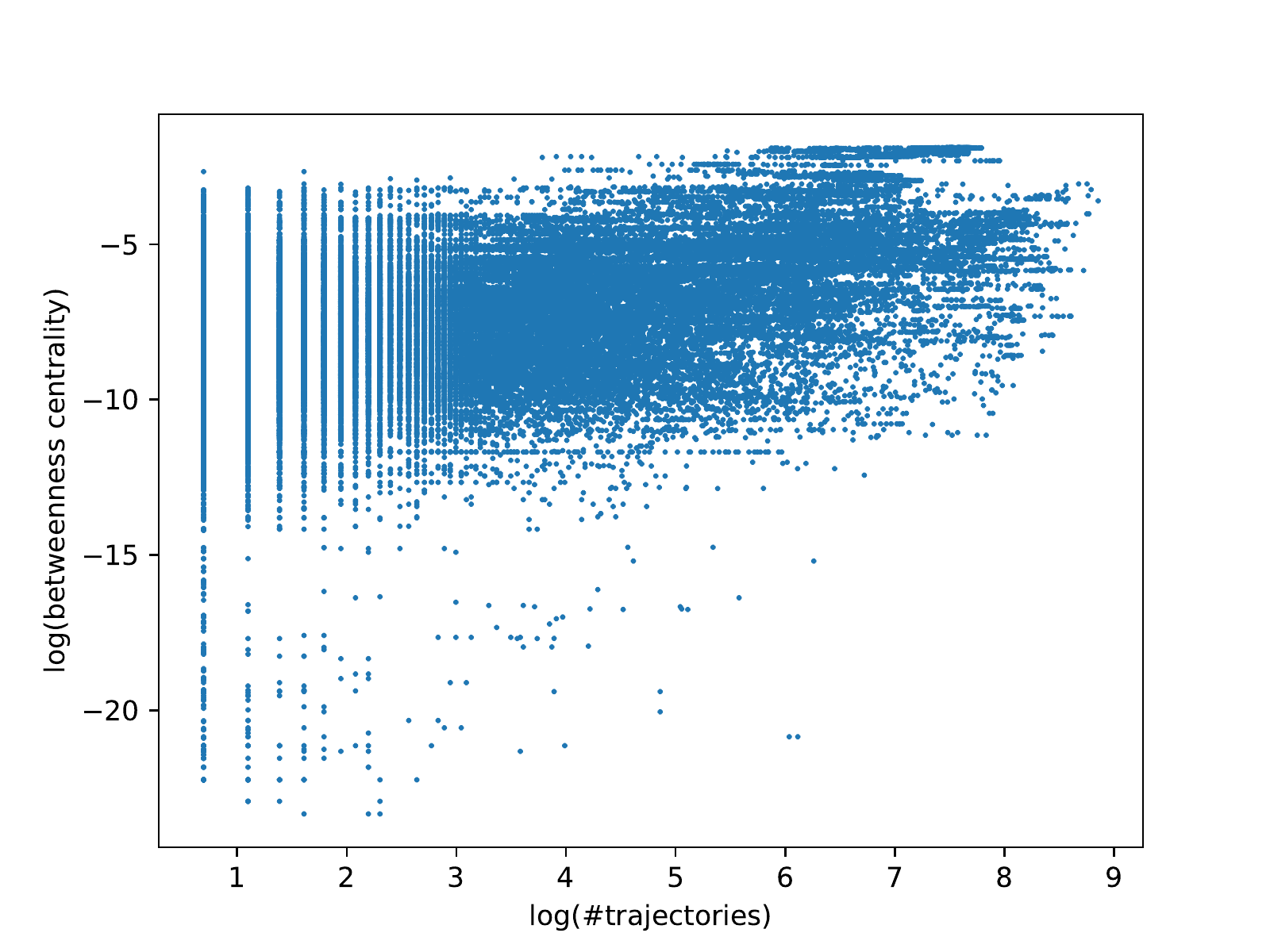}
  \caption{Scatter plot OSM node betweenness centrality vs. number of trajectories passing through each node (logarithmic scale).}
  \label{fig:freqvstraj}      
\end{figure}

Thus, therein lies a dilemma: is a segment from map $M_1$ which has not carried any trajectory a closed road segment or it simply did not see a trajectory due to its peripheral nature? To answer this dilemma we initially aimed to exploit the OSM meta-data of OSM road segments such as road type, speed limit, number of lanes or one-way tag. However, the OSM meta-data appears to be rather sparse and is unlikely to give us the relevant road importance score which would help answering the above dilemma. 

\begin{figure}
\centering
  \includegraphics[width=0.7\textwidth]{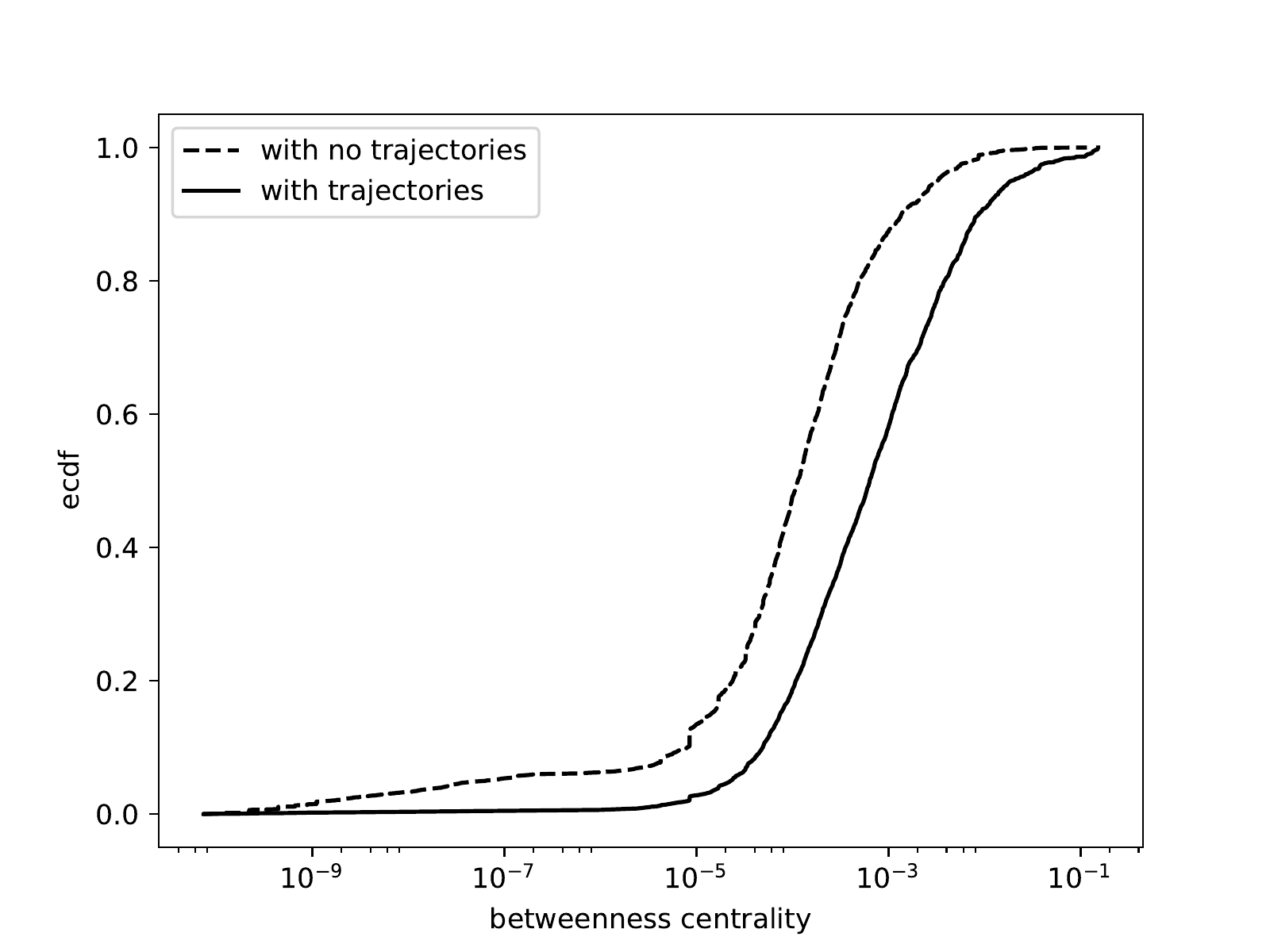}
  \caption{Empiric CDF of node centrality for two classes of nodes: those who lie on at least one trajectory and those who do not. Node betweenness centrality is generally much smaller among nodes with no trajectories.}
  \label{fig:ecdf_centrality}      
\end{figure}

We address the aforementioned question by evaluating the node betweenness centrality (BC)\footnote{We believe using another node-centrality measure would likely give similar results, though we do not evaluate the impact of the choice of centrality measure in this work. However, the use of betweenness is consistent with the problem definition in Section~\ref{sec:problem}.} in map $M_1$. The BC of a node acts as an indicator of the importance of the node in the graph $M_1$, and not-surprisingly we see a strong dependence between the centrality of a given road segment and the number of trajectories
in our data that pass through it. As seen in Figure \ref{fig:freqvstraj}, the trend is that the more trajectories a node has the higher BC and vice versa. In Figure \ref{fig:ecdf_centrality} we depict the empiric CDF of node BC for two classes of nodes: those who lie on at least one trajectory and those who do not. We observe that BC mean/median among the nodes which lie on at least one trajectory is an order of magnitude larger than among the nodes which are not carry any trajectory. 

Based on these observations, we declare the road segment closed if it has no trajectories passing by it and its BC is greater than the threshold $\gamma$. We choose $\gamma = 0.01$ to shave off the tail of the BC distribution among the nodes with no trajectories. Such $\gamma$ identifies a handful  of roads which are closed which we confirm by inspecting each one of them. In addition to those closed roads which are a sequence of closed nodes (with $BC > \gamma$) there are several nodes which are candidates for closure but are isolated from the other candidates. In order to declare the road closed we require that at least 100 meters segment (approximately 5-6 nodes) with corresponding nodes to be candidates for closure. 

We would also like to point out that the proposed methodology allows us to infer inconsistencies between the OSM data and the traffic reality as captured by the GPS data. Namely, several roundabouts (formed by nodes with $BC > \gamma$) are represented in OSM as two way streets, however the clock-wise direction in those roundabouts is not matched by any trajectory and hence it is correctly identified as closed road (in that direction) which is an unexpected benefit of using the method described above. 

\subsection{Road Closure as Anomaly Detection}
\label{sec:anomalydetection}
The method described in the previous section detects the road closures which have happened before the data collection started and it is applicable only to major roads - those with high betweenness centrality. However, for roads which get closed during the data collection we develop an anomaly detection module which monitors the traffic on each road segment and identifies ``abnormal" gaps in the traffic stream.  

For each node in the map $M_1$ we track the list of timestamps each time a trajectory is matched to that node. Note that sometimes a trajectory may have multiple records which are mapped to the same node (e.g., if the node is near a traffic light and the vehicle is static it will generate multiple data records which map to the same node in the map) and hence we only record the first match of the trajectory at the node and ignore the others. 

\begin{figure}
\centering
\includegraphics[width=0.7\textwidth]{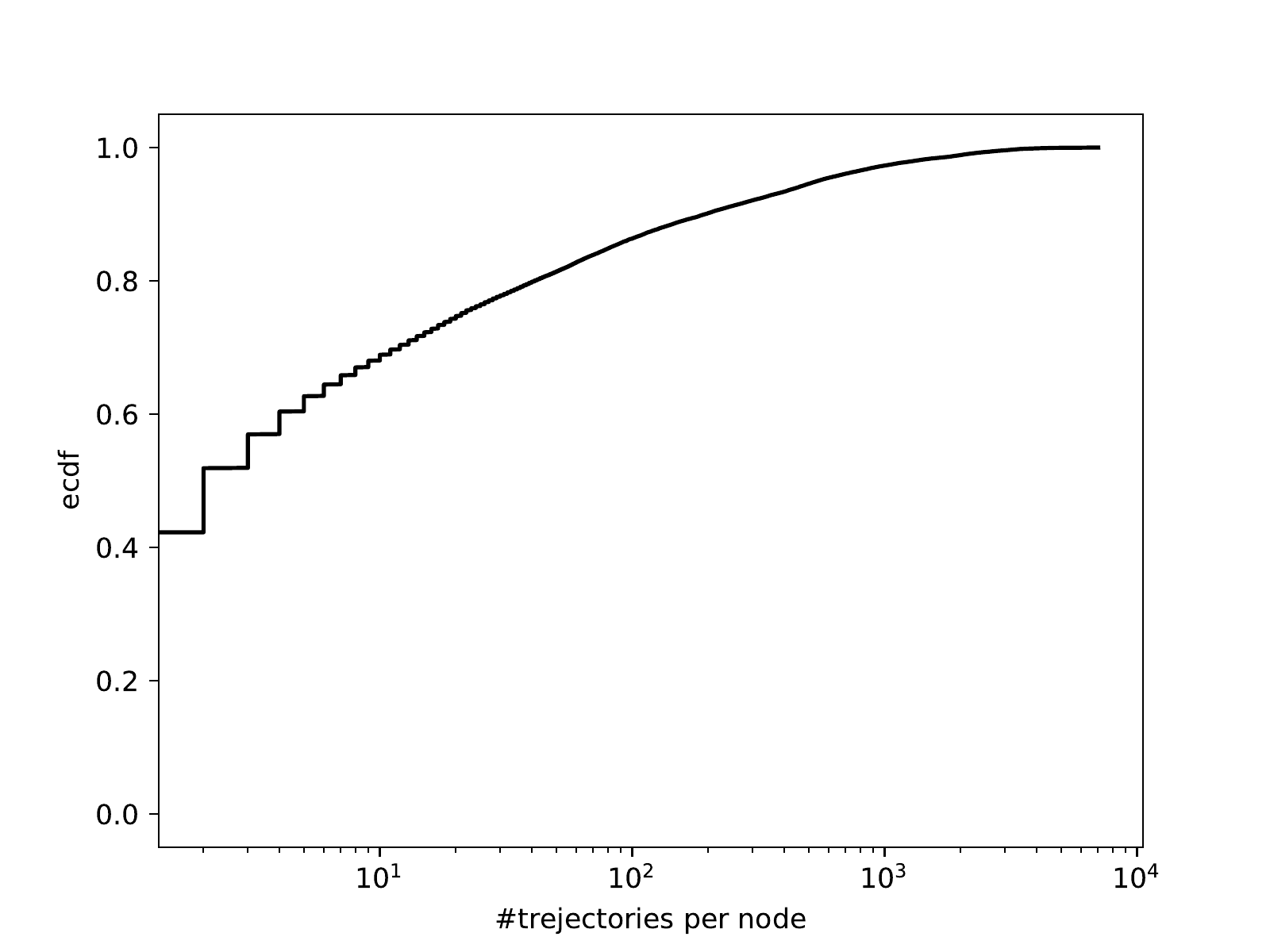}
\caption{Empiric CDF of the number of trajectories per node for all nodes in the OSM map. In our dataset only 18\% of nodes have more than one trajectory per day in average.}
\label{fig:ecdf_trajectpernode}      
\end{figure}

As described previously, the road popularity (measured by number of trajectories which pass by it) distribution is rather skewed. In Figure \ref{fig:ecdf_trajectpernode} we plot the number of trajectories that are mapped to every OSM node in our dataset and observe that a large fraction of nodes have only a handful of trajectories which pass by it. Consequently, detecting anomalies on such low-frequency roads is rather challenging. 

To detect the road closure during the data collection, each node $v$ in the OSM graph maintains $mean_v(t)$: the \textit{average} inter-arrival time among all trajectories which have passed that node until time $t$. In addition to that it also maintains the time \textit{elapsed} since the last trajectory: $e_v(t)$. Note that for optimization reasons, the time elapsed is also computed when needed such as a case where a route query is triggered.

We declare the node closed at time $t$ if:
$$e_v(t) > \alpha \cdot mean_v(t)$$
where $\alpha$ is a parameter which determines how conservative we are when deciding to declare the road closed. Small values of $\alpha$ may declare roads closed prematurely, while with large $\alpha$ it may take a long time before a closed road is declared as such. 

\begin{figure}
\centering
\includegraphics[width=0.7\textwidth]{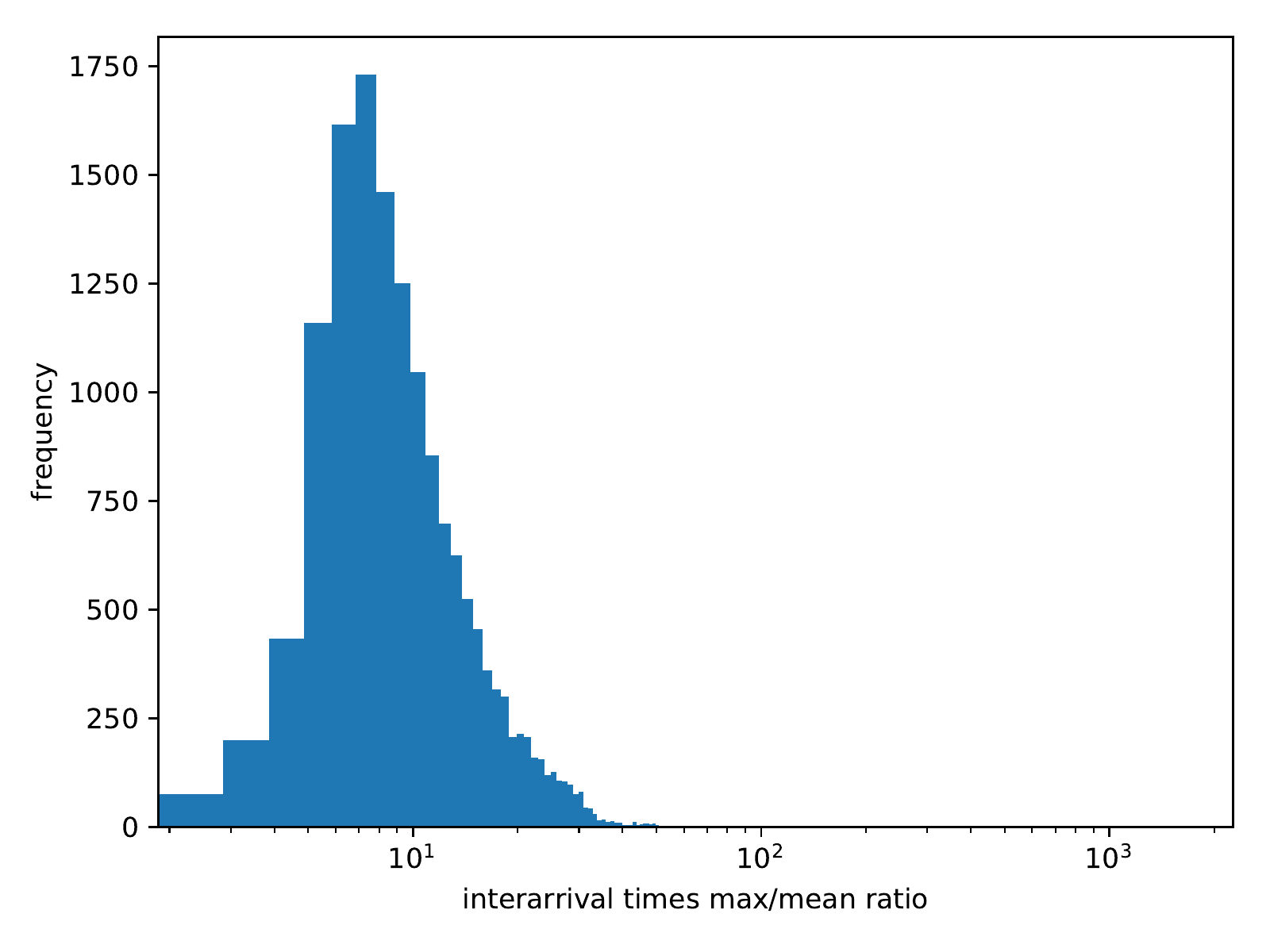}
\caption{The distribution of the ratio between maximum and average inter-arrival times for all nodes with at least 2 trajectory per day (in average). Most of the distribution falls in the range 1-40 with outliers corresponding to the nodes depicted in Figure \ref{fig:closedroadrayan}}
\label{fig:max-meanratio}      
\end{figure}

\begin{figure}
\centering
  \includegraphics[width=0.7\textwidth]{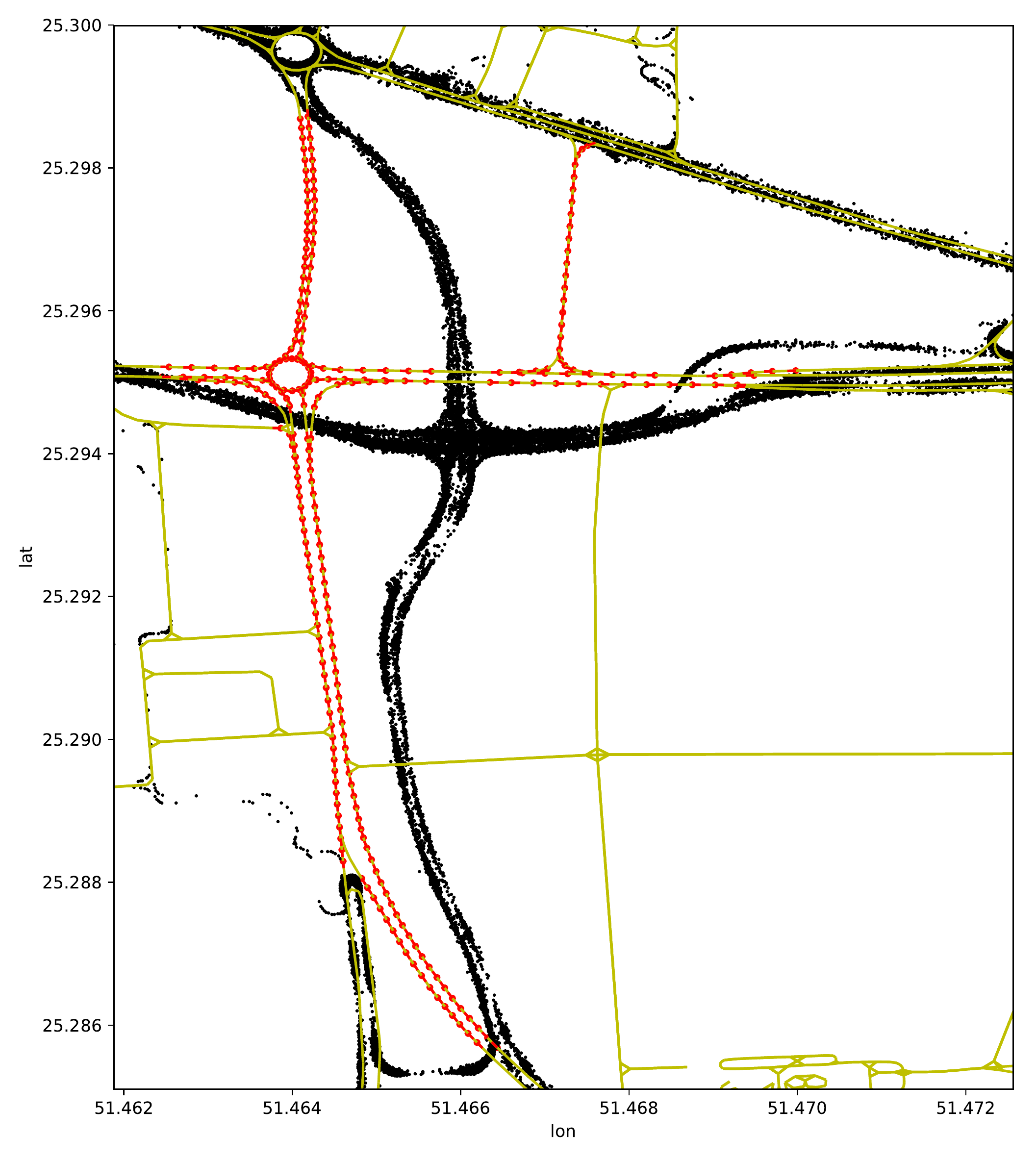}
  \caption{Detected closed OSM road segments (red). OSM road network (yellow). GPS points after the road closure (black).}
  \label{fig:closedroadrayan}      
\end{figure}

To understand what is the right choice of $\alpha$ in Figure \ref{fig:max-meanratio} we depict the histogram of the ratio between the maximum and the average trajectory inter-arrival time for all nodes which receive at least 2 trajectories per day, in average. We observe that the distribution of the max-to-average ratio is rather wide, and there is not clear cut-off point.  However, most of the distribution is in the range between 1 and 40 with only a few nodes with the ratio greater than 40. Hence we choose $\alpha = 40$. Such choice results in only one closed road-section depicted in Figure \ref{fig:closedroadrayan} during our 2-month long observation. It involves a closed roundabout and respective access roads. 

Finally, note that choosing a smaller $\alpha$ is likely to identify temporary road closures. However, since we could not confirm whether or not such nodes correspond to actual road closure or they simply fall in the tail of the distribution we leave the detailed discussion of temporary closures to future work. 

\section{Evaluation}
\label{sec:experiments}
In this section we will exploit the GPS trajectory data to evaluate the quality of the fused map. 
     
\subsection{Data}

As we discussed earlier, our map inference process uses data generated by a fleet of vehicles with GPS-enabled devices. In this paper we utilize the datasets from Doha (Qatar) with around 400 vehicles, 11 Million GPS points (sampled every $10s$). The dataset includes all GPS data points which fall into a rectangle (in $lat,lon$ coordinates) of 6km$\times$8km in an urban region in the city of Doha with a mixture of highways, high and medium volume roads, capillary streets, and roundabouts. Every data record contains: timestamp, latitude, longitude, speed, and heading of the moving direction of the vehicle. Heading is measured in angles against the North axis in degrees reporting values from $0$  to $360^{\circ}$. 

We preprocessed the data to eliminate those data points with speed $\leq 5kmph$ which are known to have non-trivial noise when reporting location. 

\subsection{Using trajectory data to evaluate maps}
In this section we analyze how well can we match trajectories to the maps. For a map $\mathcal{M}$ and a trajectory $\tau = (p_1,\dots,p_k)$ we denote by $\delta(\tau,\mathcal{M})$ the maximum distance between the points on the trajectory $\tau$ and $\mathcal{M}$:
$$\delta(\tau,\mathcal{M}) = \max_{p_i\in\tau}\min_{(u,v)\in \mathcal{M}}v(p_i,(u,v))$$
where $v(p_i,(u,v))$ is simple distance to line segment in geo-distance, measured in meters. 

In our data we split all the trajectories in two subsets: training and test. We use the training set for constructing map $\mathcal{M}_2$ and the test set of trajectories for evaluating the matching distance. Since many trajectories from the same driver coincide, we make sure that trajectories from the same driver do not fall into both training and testing data. To that end, we split the set of drivers into training/test drivers (75\%/25\% split) and assign all the trajectories from the training/test driver into training/test trajectory dataset, respectively. 

For automatic map inference we use \kha \cite{stanojevic2017kharita}, but note that using any other automatically inferred map \cite{BiagioniE12,cao2009gps,chen2016city,edelkamp2003route} could be used with relatively small (small, since only a handful of roads are being added to the map) impact on the final fused map.

For each trajectory in the test data we evaluate $\delta(\tau,\mathcal{M}_1)$, $\delta(\tau,\mathcal{M}_2)$, and $\delta(\tau,\mathcal{M}_1\bigoplus \mathcal{M}_2)$, where $\mathcal{M}_1$ is the underlying (OSM) map, $\mathcal{M}_2$ is the automatically inferred map using the training trajectory data and $\mathcal{M}_1\bigoplus \mathcal{M}_2$ is the merged map.

\setlength\tabcolsep{1em}
\begin{table}
\centering
\begin{tabular}{l rrr}
\toprule
$\delta(\cdot,\cdot)$ & mean & median & 99th-\% \\ 
\midrule
OSM & 40.3m & 9.3m & 333m \\ 
automatic & 12.3m & 9.1m & 70.4m \\ 
merged & 8.1m & 6.0m & 53.4m \\ 
\bottomrule
\end{tabular} 
\vspace{3mm}
  \caption{Trajectory matching distance.}
  \label{tab:trajectmatching}      
\end{table}

In Table~\ref{tab:trajectmatching} we report the mean, median and 99th-percentile trajectory matching distance for the three maps. All three metrics (mean, median and 99th-percentile) are minimized for the merged map and are around one third smaller than for automatic map. The improvements in trajectory matching come for two reasons. On one hand, trajectories which follow the new roads non-existing in the OSM map, but discovered by the automatic map, enjoy better matching in the merged map. On the other, the parts of the trajectories which correspond to the roads which are not covered in the training data, are likely to be covered in the OSM map and hence in the merged map.

\section{Conclusion}
\label{sec:conclusion}

In this paper, we proposed a new map update paradigm: map fusion. Instead using a customized map-inference algorithm when updating a map, we allow any map to be fused to the underlying (say OSM) map. Such fusion allows for quick map updates, with minimal changes to the high-quality underlying map. In addition to the map fusion, we also study in detail the road closure detection and propose two methods which efficiently detect road closure by comparing the statistical expectation of the traffic on a road segment against the actual traffic. 
\bibliographystyle{spbasic}
\bibliography{mapinf} 

\end{document}